\documentclass[journal=jacsat,manuscript=article]{achemso}

\usepackage{graphicx} % Required for inserting images

\usepackage{xr}

\makeatletter
\newcommand*{\addFileDependency}[1]{% argument=file name and extension
  \typeout{(#1)}
  \@addtofilelist{#1}
  \IfFileExists{#1}{}{\typeout{No file #1.}}
}
\makeatother

\newcommand*{\myexternaldocument}[1]{%
    \externaldocument{#1}%
    \addFileDependency{#1.tex}%
    \addFileDependency{#1.aux}%
}

\myexternaldocument{SI}

\usepackage{hyperref}
\hypersetup{
    colorlinks=true,
    linkcolor=blue,
    filecolor=magenta,      
    urlcolor=cyan,
    pdftitle={Overleaf Example},
    pdfpagemode=FullScreen,
    }
    
\author{Anika J. Friedman}
\affiliation{Department of Chemical and Biological Engineering, University of Colorado Boulder, Boulder, CO 80303}
\author{Wei-Tse Hsu}
\affiliation{Department of Chemical and Biological Engineering, University of Colorado Boulder, Boulder, CO 80303}
\author{Michael R. Shirts}
\affiliation{Department of Chemical and Biological Engineering, University of Colorado Boulder, Boulder, CO 80303}
\email{michael.shirts@colorado.edu}

\title{Multiple Topology Replica Exchange of Expanded Ensembles (MT-REXEE) for Multidimensional Alchemical Calculations}

\begin{document}

\maketitle
\begin{abstract}
    Relative free energy calculations are now widely used in academia and industry, but the accuracy is often limited by poor sampling of the complexes conformational ensemble. To address this, we have developed a novel method termed Multi-Topology Replica Exchange of Expanded Ensembles (MT-REXEE). This method enables parallel expanded ensemble calculations, facilitating iterative relative free energy computations while allowing conformational exchange between parallel transformations. These iterative transformations are adaptable to any set of systems with a common backbone or central substructure. We demonstrate that the MT-REXEE method maintains thermodynamic cycle closure to the same extent as standard expanded ensemble for both solvation free energy and relative binding free energy. The transformations tested involve simple systems that incorporate diverse heavy atoms and multi-site perturbations of a small molecule core resembling multi-site $\lambda$ dynamics, without necessitating modifications to the MD code, which in our initial implementation is GROMACS. We outline a systematic approach for topology set-up and provide instructions on how to perform inter-replicate coordinate modifications. This work shows that MT-REEXE can be used to perform accurate and reproducible free energy estimates and prompts expansion to more complex test systems and other molecular dynamics simulation infrastructures.
\end{abstract}

\section{Introduction}
%General introduction and present the problem we aim to help solve
Alchemical free energy calculations have been an important tool for accurately ranking the free energy difference between two states, but more recently these methods have become more common and widely used due primarily to improvement in the accuracy and computational efficiency of these calculations~\cite{chodera_alchemical_2011, song_evolution_2020, york_modern_2023, shirts_chapter_2007}. The primary application of alchemical free energy calculations is in the lead optimization stage of drug development~\cite{abel_critical_2017, lee_alchemical_2020, jorgensen_efficient_2009, cournia_free_2021, williams-noonan_free_2018}. Many leading pharmaceutical companies use alchemical relative binding free energy calculations (RBFE) to discover compounds with higher binding affinity to their target protein~\cite{muegge_recent_2023, mobley_perspective_2012, homeyer_binding_2014}. The introduction of alchemical RBFE calculations to the workflow for these companies has been demonstrated to save time and money in the development of new drug compounds; however, there are still numerous limitations to the wide-spread application of these techniques~\cite{cournia_free_2021}. Compared to other computational methods which have gained popularity in recent years like deriving free energy estimates from Markov state models, machine learning predictions, or end point methods (ie. MM/GBSA), alchemical free energy calculations are often times significantly slower which is balanced by their increased accuracy~\cite{ngo_benchmark_2021, cournia_relative_2017, luo_challenges_2019, hata_binding_2021, wang_end-point_2019}. The accuracy of alchemical free energy calculations is somewhat limited by the accuracy of molecular mechanics force fields and the interplay between protein and small molecule force fields as well as handling charge changing mutations, which are being addressed well by other efforts~\cite{lin_force_2019, lim_benchmark_2020}, but insufficient conformational sampling due to limitations in available computational resources is often a primary factor. Numerous enhanced sampling methods have been developed to minimize simulation times necessary for the calculation of sufficiently accurate free energy estimates~\cite{jiang_enhanced_2023, de_ruiter_advances_2020}. Despite the progress made in recent years, there are still many systems for which RFE techniques are not normally applied due to the presence of flexible binding interfaces, which significantly increase the challenge of computing and accurate free energy estimates.

%Introduce different enhanced sampling methods and MT-REXEE
We are certainly not the first to try to resolve or improve the issue of conformational sampling in alchemical RBFE calculations and there are far too many methods to review here, but we highlight a few methods which are similar to what we propose~\cite{lee_aces_2023, azimi_relative_2022, zhang_alchemical_2024, wu_alchemical_2021, perthold_accelerated_2018, xu_accelerating_2021, konig_alternative_2020, hahn_overcoming_2020, bussi_using_2020, lagardere_lambda-abf_2024}. Methods like Hamiltonian replica exchange (HREX) and expanded ensemble (EE) use the alchemical $\lambda$ variable to their advantage by either allowing conformational swaps between $\lambda$ intermediates as is the case for HREX or by allowing the $\lambda$ states to change while retaining the conformation from the previous state~\cite{procacci_does_2022, zhang_expanded_2021, jiang_reduced_2018, wan_accuracy_2020}. Alchemical metadynamics allows perturbation along both the alchemical as well as configurational collective variables to enhance configurational sampling in alchemical free energy calculations~\cite{hsu_alchemical_2023}. This method does require prior knowledge of the slowest degrees of freedom to select your collective variable. $\lambda$-dynamics is an enhanced sampling method which allows for the evolution of $\lambda$ along the intrinsic free energy landscape rather than utilizing pre-determined $\lambda$ state spacing used by the other methods discussed previously~\cite{kong_dynamics_1996, hayes_how_2024}. 

\begin{figure}
    \centering
    \includegraphics[width=0.9\linewidth]{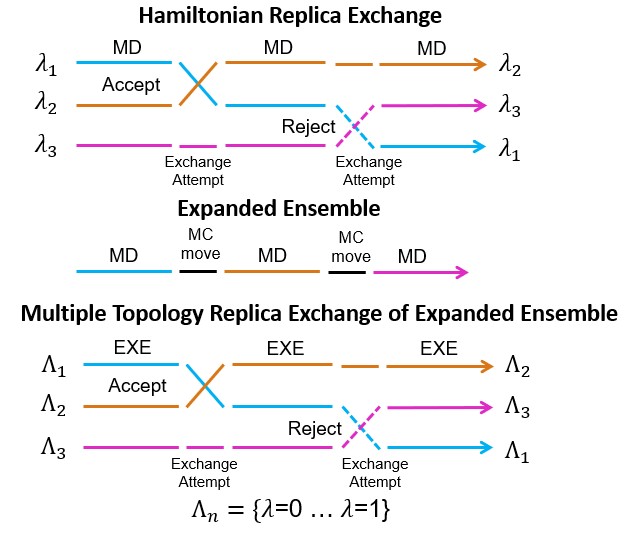}
    \caption{Traditional Hamiltonian replica exchange involves running several parallel simulations which each have a constant $\lambda$ value which varies from 0 to 1. EXE calculations require only a single simulation which moves between adjacent lambda states at a pre-determined interval. MT-REXEE combines these methods to allow multiple parallel EXE simulations to run and swap coordinates between adjacent replicas.}
    \label{fig:MT-REXEE}
\end{figure}

Our group previously developed replica exchange of expanded ensemble (REXEE), which builds off the principles of HREX and EXE in order to sample sub-sets of $\lambda$ intermediate states to enhance sampling of single relative transformations~\cite{hsu_replica_2024}. The multiple topology replica exchange of expanded ensemble (MT-REXEE) method further extends the REXEE method to couple several different alchemical transformations together to increase conformational swapping while computing the relative free energies of these transformations (Figure \ref{fig:MT-REXEE}). This leverages the $\lambda$ intermediate states to increase conformational sampling within each parallel transformation as well as sharing conformational sampling between other connected transformations.

%How is MT-REXEE different from MSlD
The idea of increasing conformational sampling by doing many simultaneous transformations is not unique either, but our method approaches this implementation is a distinct manner. Many methods have been developed to implement multi-site $\lambda$-dynamics (MS$\lambda$D) which allows for sampling of multiple functional groups at multiple sites on a shared ligand core to be sampled within a single simulation~\cite{hayes_blade_2021, robo_fast_2023, raman_automated_2020, hu_comprehensive_2024}. MS$\lambda$D has been demonstrated to produce accurate binding free energy estimates for relatively inflexible protein-ligand complexes faster then traditional AFE methods~\cite{knight_multisite_2011, ding_cdocker_2018, champion_accelerating_2023}. However, since these methods only sample a single protein conformation while the numerous ligands are sampled, the increased sampling is only on the part of the ligand. This becomes relevant for more flexible binding pockets especially with those binding pockets which may under ligand-induced conformational changes. These methods are also limited to the CHARMM and OpenMM simulation engines. MT-REXEE samples separate protein and ligand conformations in each transformation and is implemented as a wrapper for the simulation engine which means it can be applied to any engine with minor adjustments. MS$\lambda$D methods will generally  be computationally less expensive if the conformational sampling is sufficient in the protein pocket. Likely, MT-REXEE will be more successful as a method specifically for instances in which MS$\lambda$D or other less expensive methods fail, such as the instance of large flexible ligands in flexible binding pockets, and is also much more amenable to generalizations beyond the GROMACS implementation described in this paper.

%Generalize to non-drug discovery applications
Though the primary application of AFE methods is for small molecule binding free energy calculations for drug discovery, this is certainly not the only potential application. The increase in conformational sampling on behalf of both binding partners lends itself to more diverse problems like binding free energy of protein-peptide complexes for the development of peptide drugs as well as protein-protein binding complexes. 

%Loop back around to the method and how we demonstrate that it works
In introducing the MT-REXEE method, we demonstrate first that the introduction of these conformational swaps between different alchemical transformations does not introduce additional error into either relative solvation free energy (RSFE) or RBFE calculations beyond existing methods. To demonstrate this we utilize two relatively simple test systems. First we compute the RSFE of a set of aldehydes with varying numbers of carbons (4--16). We also look at RSFE and RBFE of MUP1 inhibitors which are a commonly used benchmarking target for AFE methods~\cite{robo_fast_2023}. By demonstrating that we can achieve similar results to running stand-alone EXE simulations, we show that this method can then be applied to more complicated systems in which we can see a benefit from our increased conformational sampling.

\section{Methods}
\subsection{Running Free Energy Simulations}
\subsubsection{System Set-up}
In this study we utilize two test systems: the solvation of aldehydes and protein-thiazole binding. The initial conformations for the solvation free energy of the aldhyde systems (C\textsubscript{4}OH\textsubscript{8}-C\textsubscript{16}OH\textsubscript{32}) were constructed using Avogadro~\cite{hanwell_avogadro_2012} and the initial conformations for both the solvation free energy and binding free energy were derived from the crystal structure (PDB 1IO6). All small molecules were parameterized using GAFF2~\cite{he_fast_2020} and the protein was parameterized using Amber \textit{ff14sb} force field\cite{maier_ff14sb_2015}. The relative topologies for all systems were created using PMX~\cite{gapsys_pmx_2017} followed by post-processing to remove all dihedral energies between dummy and real atoms within the hybrid topologies. For vacuum simulations, the simulation box was constructed to maintain a 3 nm distance between the molecule and the periodic boundary. The box for all simulations conducted in solvent were constructed to maintain a minimum distance of 1 nm between the molecule and the periodic boundary and then filled with TIP3P solvent.

A standard procedure was used for equilibration all systems. This procedure consisted of an initial energy minimization to a threshold of 50 kJ/mol$\cdot$K followed by a 100 ps NVT equilibration using the Bussi-Parrinello thermostat~\cite{bussi_canonical_2007} to maintain a temperature of 300K. A 100 ps NPT equilibration was also performed with the stochastic cell rescaling barostat~\cite{bernetti_pressure_2020} for all calculations performed in solvent to maintain a pressure or 1 atm. All simulations were performed with a 2 fs timestep. Additional simulation set-up information can be found for each simulation in the \href{https://github.com/ajfriedman22/MT-REXEE/tree/main}{GitHub} repository.

\subsubsection{Validation Free Energy Simulations}
Absolute solvation free energy (AFE) calculations were performed in order to identify any error associated with the relative free energy calculation performed with both EXE and MT-REXEE. The procedure for running both absolute and relative free energy simulations was identical other than the total simulation run time which was 40 ns for relative vacuum simulations, 60 ns for relative solvent simulations, and 100 ns for absolute solvation free energy and relative protein complex simulations. Free energy estimates were computed using MBAR~\cite{chodera_replica_2011}. The $\lambda$-state spacing was determined for AFE of both systems independently in order to maintain sufficient off-diagonal state overlap of at least 0.1. The production simulations were run using the expanded ensemble formalism~\cite{lyubartsev_new_1992,chodera_replica_2011}. The Hamiltonian state was changed every 100 time steps (or 200 fs) and moves were made using the Metropolis-Gibbs algorithm~\cite{liu_peskuns_1996}. The state weights were updated using the Wang-Landau algorithm~\cite{belardinelli_wang-landau_2007} and were fixed  when the Wang-Landau incrementor was less than the 0.001 threshold. All settings for the expended ensemble simulations can be found for each simulations in the \href{https://github.com/ajfriedman22/MT-REXEE}{MT-REXEE repository}.

\subsubsection{Coordinate Modification}
A generalized method for coordinate modification is implemented in the REXEE code which can be found on \href{https://github.com/ajfriedman22/MT-REXEE/blob/main/code/swap_gro.py}{GitHub}. First, we determine which $\lambda$ state is being sampled in each parallel simulation. It two simulations are at compatible end states in the final frame of the previous simulation then a swap is attempted. Which simulations can swap are determined by the swapping pattern, but the two simulations must share a common molecular endpoint of in each of their transformations . For example, simulation 1 samples the relative transformation from molecule A to B and simulation two samples the relative transformation from molecule B to C. If simulation 1 is sampling $\lambda=1$ in the last frame of the trajectory and simulation 2 is sampling $\lambda=0$ then a swap will be performed. Prior to swapping coordinates, any discontinuities in the molecule due to periodic boundary conditions were resolved by shifting atoms across the periodic boundary to prevent alignment issues. 

The molecules are then aligned to enable building in coordinates for dummy atoms which differ between the two adjacent replicas. The alignment occurs across the bond in which a different functional group must be substituted and therefore can occur multiple times depending on the differences in the structure of the dummy atoms. An anchor atom which is both present and real in the $\lambda$ state being sampled and an alignment atom which is the real atom that becomes a dummy atom when the functional group being added becomes real are selected based on connectivity within the molecule. First, a translation is performed to align the anchor atom in both molecules. Then the alignment atom is present in the pre-swap molecule is rotated along an axis perpendicular to the vectors between the anchor atom and the alignment in both molecules until the alignment atom in the pre-swap molecule reaches the position in the reference molecule. Finally, the missing atom which forms a covalent bond with the anchor atom is rotated along the vector formed between the anchor atom and the alignment atom to match the angle formed between a common atom, the anchor, and the missing atom is consistent with the pre-swap configuration. 

The coordinates for all atoms including the constructed atoms are then written to a GRO file with 7 decimal place precision. The standard GRO file contains 3 decimal places, but we found that the additional precision is necessary to ensure that the systems energies match before and after swaps occur. In order to achieve additional precision in the input coordinates the coordinate modification file reads in a non-compressed trajectory (.trr format) as GROMACS can read additional precision in GRO files, but does not output higher precision in this format. There are no issues with clashing water molecules or protein residues, as the atoms which have been constructed from this coordinate modification begin the simulation as fully non-interacting dummy atoms which are incapable of clashing with real atoms and the system is allowed to equilibrate before these interactions are restored. The swapping algorithm runs on a single core and takes 0.3--0.8 seconds depending on the size of the system and whether periodic boundary breaks need to be corrected. 

The introduction of the coordinate modification function thus has a small contribution to the run time of the full algorithm. The wall-time for a single 20 ps iteration for a vacuum, solvent, and complex simulation are $11.1 \pm 0.2$, $12.3 \pm 0.2$, and $28.5 \pm 0.4$ seconds which are all significantly larger (14-57x) than the 0.3-0.8 seconds introduced per conformational swatch meaning that introducing more complex program structure to performing conformational swap itself is not necessary. For a total simulation length of 20 ns for vacuum simulations the coordinate modification leads to a 1.5\% increase in total computational cost. The coordinate modification step scales with the degree of modification rather than the total number of atoms in the system so the increase in computational cost decreases with system size for an increase in computational cost of 0.8\% and 0.2\% increase for the solvent and complex simulations respectively.

\begin{figure}[h]
    \centering
    \includegraphics[width=0.9\linewidth]{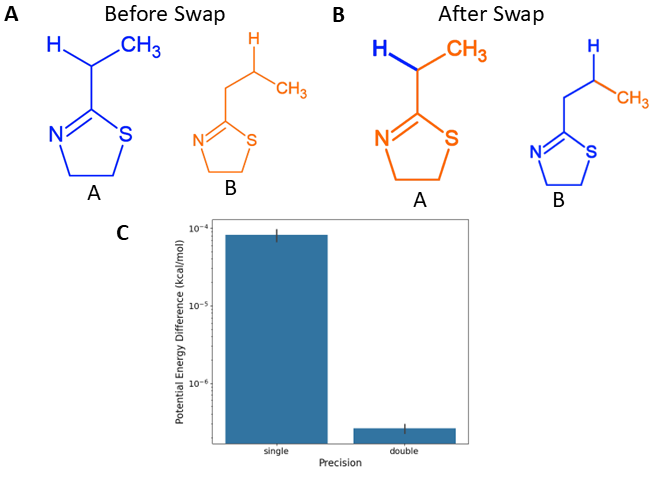}
    \caption{Here we diagram the origin of the coordinates in the final configuration that is used for the next simulation iteration. The explicitly shown hydrogens are dummy hydrogens which are additional to the hydrogens that would usually fully satisfy the bonds for that heavy atom. The potential energy for bonded interactions in blue should match the potential energy for bonded interactions in molecule A before the swap was performed and those interactions in orange should match those from molecule B.}
    \label{SI:coord_swap_ener}
\end{figure}

The coordinate modification step must conserve the potential energy of the two segments of the molecule which are run in separate simulations, i.e. the $\lambda=0$ state of one simulation and the $\lambda=1$ state of a companion simulation. 
If we are swapping coordinates between molecule A and molecule B as shown in figure \ref{SI:coord_swap_ener}A, we need to reconstruct coordinates for the dummy hydrogen for the new configuration of molecule A and coordinates for the methyl group for the new configuration of molecule B. The new configuration for A would thus have a subset of atoms which should have bonded parameters corresponding to the potential energy for the initial configuration in A and a subset corresponding to the initial configuration in B shown by molecule color in figure \ref{SI:coord_swap_ener}B. 

We computed the potential energy of bonds, angles, torsions, and van der Waals interactions of the small molecule before and after the conformational swap. The interactions were then separated based on the origin molecule for which the interactions should match (ie. energies corresponding to blue atoms before the swap should match blue atoms after the swap based on figure \ref{SI:coord_swap_ener}). We computed these energies for three replicates of the acyl system in both vacuum and solvent as well as the MUP1 ligand system in vacuum and solvent. On average the total deviation in potential energy with this coordinate swapping method was found to be $7.1 \times 10^{-5} \pm 8.6 \times 10^{-5}$ kcal/mol (Figure \ref{SI:coord_swap_ener}C). This error can be decreased to $2.4 \times 10^{-7} \pm 2.5 \times 10^{-7}$ kcal/mol by running GROMACS in double precision and increasing the number of decimal places in the GRO file from 7 to 10 which leads us to conclude that the dominant deviation in potential energy is due to the precision in the coordinates and not a systematic error being introduced.  We note that the released code includes a ``paranoid mode'' which slows down exchange but allows potential energy validation of each switch.

\subsubsection{MT-REXEE Protocol}
Each MT-REXEE simulation was conducted using the same $\lambda$ state spacing as their corresponding EXE simulation and was run for the same total duration (following weight equilibration). We included redundant end states ($\lambda = 0$ and $\lambda = 1$) for these simulations in order to increase the frequency at which swaps were able to occur. We added these redundant end states such the 1/3 of total number of states were at $\lambda = 0$, 1/3 were at $\lambda = 1$, and 1/3 were sampling intermediate $\lambda$ states. All $\lambda$ states were initiated with 0 weights to match the EXE procedure and weights were equilibrated using the MT-REXEE method. For both solvation and binding free energy of both the aldehyde and thiazole systems the conformational swaps were attempted every 20 ps unless otherwise stated. The 20 ps value was determined by running MT-REXEE simulations in vacuum and solvent for the MUP1 ligand systems using different swapping rates (10, 20, 50, and 100 ps). We found there was no significant difference in the free energy estimates while varying the swapping rate, but the round trip time (time to sample all states) significantly decreased as the swapping frequency decreased until a frequency of 20 ps (Figure~\ref{SI:equil_and_opt}C--D).  This inclusion of redundant states reduces the total fraction of intermediate time spent at intermediate states compared to EXE simulations, but as demonstrated in the results section, significantly increases the exchange rate between states.  The exact trade-off between these two effects will be system dependent and we do not attempt in this study to generalize the determination of this time scale to other systems. 

\subsection{Analyzing Free Energy Simulations}
\subsubsection{Determining FE from Simulations}
For all simulations conducted in this study the free energy estimate was computed using the alchemlyb~\cite{beckstein_alchemistryalchemlyb_2024} wrapper around pymbar~\cite{shirts_statistically_2008}. Only the trajectory frames after the $\lambda$ weights were equilibrated according to the Wang-Landau iterator criteria were used to compute the free energy. The number of uncorrelated samples were determined using the alchemlyb statistical inefficiency function. Three replicates were used for each simulation and the free energy values reported are the mean of the three replicate simulations. The uncertainties reported are the standard error of the mean using the three independent replicas.

\subsubsection{Evaluating Cycle Closure}
Relative free energy calculations rely on the closure of the thermodynamic cycle for each relative transformation. We wish to evaluate whether our introduction of conformational swaps between different relative transformations affects the accuracy of our relative free energy estimates. 
\begin{figure}
    \centering
    \includegraphics[width=\textwidth]{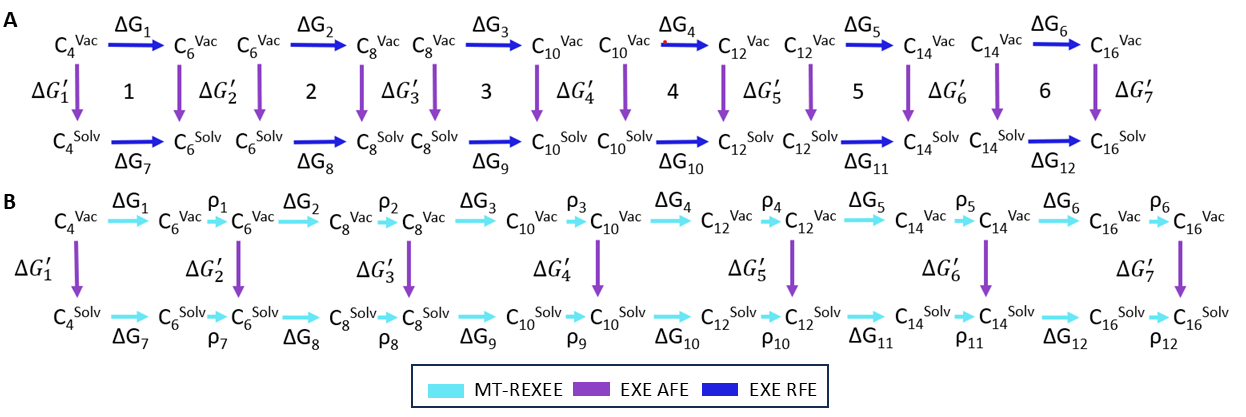}
    \caption{A) The individual thermodynamic cycles for the calculation of $\Delta\Delta$G of the iterative transformation from C4--C16 using expanded ensemble simulations. The purple absolute solvation free energy transformations and the blue relative solvation free energy transformations create independent thermodynamics cycles for each transformation. B) The thermodynamic cycle for the calculation of $\Delta\Delta$G for the iterative transformation from C4--C16 using a single MT-REXEE simulation with 6 replicas. Notably, the light blue relative solvation free energy transformations are connected through the conformational swaps between simulation instances.}
    \label{fig:thermo}
\end{figure}

Figure \ref{fig:thermo}A shows the thermodynamics cycles for each of the individual transformations for the solvation free energy for the aldehyde system. The path independence of free energies gives us equation \ref{eq:thermo1}, and from this we can get equation \ref{eq:thermo2}.
\begin{equation}\label{eq:thermo1}
    \Delta G_1 + \Delta G_B - \Delta G_7 - \Delta G_A = 0
\end{equation}
\begin{equation}\label{eq:thermo2}
    \Delta \Delta G = \Delta G_B-\Delta G_A = \Delta G_7 - \Delta G_1
\end{equation}
Equation \ref{eq:thermo2} assumes perfect matching between toplogies, but there is potentially error associated with each individual simulations such that equation \ref{eq:thermo1} becomes \ref{eq:thermo3} where $\delta_1$ is the sum of the error associated with all four calculations.
\begin{equation}\label{eq:thermo3}
    \textrm{Cycle Error} = \mid\Delta G_1 + \Delta G_B - \Delta G_7 - \Delta G_A\mid = \mid \delta_1 \mid
\end{equation}
In the MT-REXEE method there is an additional source of error that does not apply to the individual EXE calculations from the conformational swapping step. This additional error associated with the conformational swaps in the MT-REXEE method is given by $\rho_n$ in figure \ref{fig:thermo}B. This error likely comes from two sources: the treatment of dummy atoms and changes in the potential energy of the molecule before and after conformational swaps. Dummy atoms have no non-bonded energies because $\epsilon$,$\sigma$, and $q$ are all 0,  but dummy atoms still must contain some bonded interactions between the real and dummy atoms in order to prevent non-physical atom movements and maintain the general molecule structure~\cite{fleck_dummy_2021}. In this study all torsions involving real and dummy atoms are set to 0 at the dummy endpoints, but there are bond and angles which maintain their standard energy parameters in order to prevent non-physical structures from being created. The presence of these dummies with some nonzero non-bonded terms can produce some changes in the configurational space sampled whose effect must be evaluated~\cite{fleck_dummy_2021}. We do not expect this contribution to the transformation free energy to be zero, but we do expect it to be nearly the same between transformations in the solution, the vacuum and in complex with proteins. The swapping function minimizes the potential energy difference and we assert that this error should be nearly equivalent in both the solvent and vacuum transformations such that $|\rho_n-\rho_{n+6}| \approx 0$. To evaluate this error, we assume that the deviation between the relative free energy obtained from EXE and the value obtained from MT-REXEE can be attributed to this conformational swapping error given by $|\rho_n-\rho_{n+6}|$.
\begin{equation}\label{eq:thermo_MTREXEE}
    \mid\Delta\Delta G_{EXE} - \Delta\Delta G_{MTREXEE}\mid = \mid \rho_n - \rho_{n+6}\mid
\end{equation}
Equation \ref{eq:thermo_MTREXEE} assumes that the conformational sampling between the EXE and MT-REXEE simulations is equivalent such that they would be expected to produce the same result if not for additional error introduced by the conformational swapping step itself. This can reasonably be expected for the solvation free energy and binding free energy of the systems selected for this study, as there are no significant free energy barriers which may lead to significant differences in the conformational sampling from the two methods. This will likely not be the case for more complex systems for which MT-REXEE will achieve increased sampling of the full conformational space thus leading to differences in the relative free energy estimates despite minimal error introduced by the conformational swapping step.

\subsubsection{Evaluating Replicate Swapping}
In order to evaluate the path of transitions throughout the simulation, we construct replica transition paths starting from the initial state of each replica. These transition paths follow the state being sampled for each conformation. All simulations are initiated from $\lambda=0$ for their individual transformation. The transition path follows the conformation state, not the replica state, such that when a conformational swap occurs the conformational state is now sampling the region for the adjacent replica. These transition paths can be used to determine a round trip time or the time it takes a transition path to return to its originally sampled state. The mean round trip times were computed from all three replicate trajectories. For non-linear swapping paths (ie. sim 1 can swap with one other than sim 2), we cannot follow a simple swapping path and thus we instead consider the round trip time as the time it takes for all possible end states to be sampled (resetting each time all end states are visited and a round trip is recorded) rather than for swaps to occur between all simulations.

\section{Results}
\subsection{Acyl Chain System Validation}
\begin{figure}
    \centering
    \includegraphics[width=0.9\linewidth]{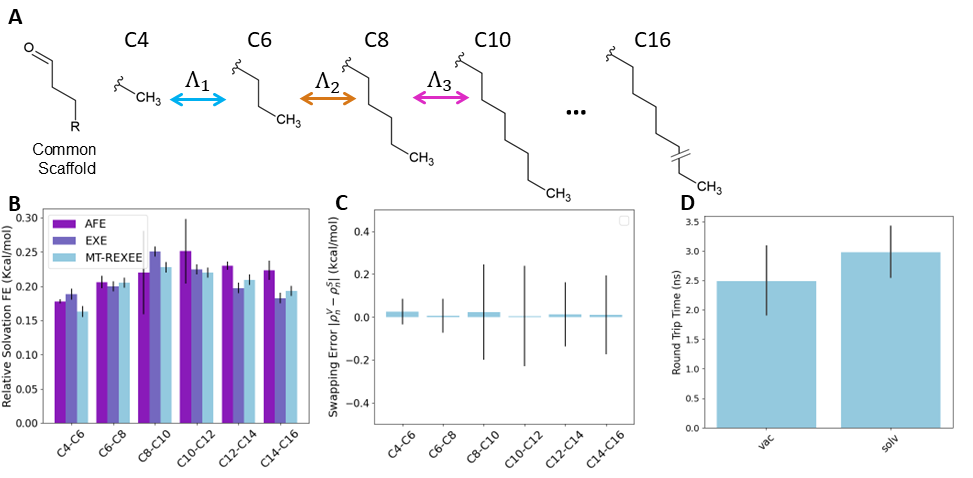}
    \caption{Validation of aldehyde chain growth test case. (A) We compute the solvation free energy of an hydrocarbon-functionalized aldehyde as we vary the number of carbons in the chain. (B) The relative solvation free energy is computed using three different methods: AFE uses the absolute solvation free energy of each end state to compute a relative value, EXE runs separate relative solvation free energy calculations for each transformation, and MT-REXEE runs parallel solvation free energy calculations with conformational swapping. (C) The error associated with the conformational swap process is statistically negligible. (D) The round trip time shows the average duration for conformational swaps, which allow shared swapping between all simulations. The rate is on the order to 2--3 ns and thus a 20 ns simulation will have on average 8 full round trips.}
    \label{fig:aldehyde_summary}
\end{figure}

Our first step in validating this enhanced sampling method is to ensure that the introduction of these conformational swaps between adjacent relative transformations does not introduce errors into the calculation. We first use a simple test system where we compute the solvation free energy for aldehyde alkane chains of varying length from 4 to 16 carbons (Figure \ref{fig:aldehyde_summary}A). This system was selected because it is relatively computationally inexpensive allowing us to ensure that we are adequately sampling the conformational space in both the EXE and MT-REXEE calculations. This allows us to isolate the contribution of conformational swapping to the free energy estimate since there should be minimal variance between individual replicas over the timescale of 20--30 ns. 

\begin{figure}
    \centering
    \includegraphics[width=0.9\linewidth]{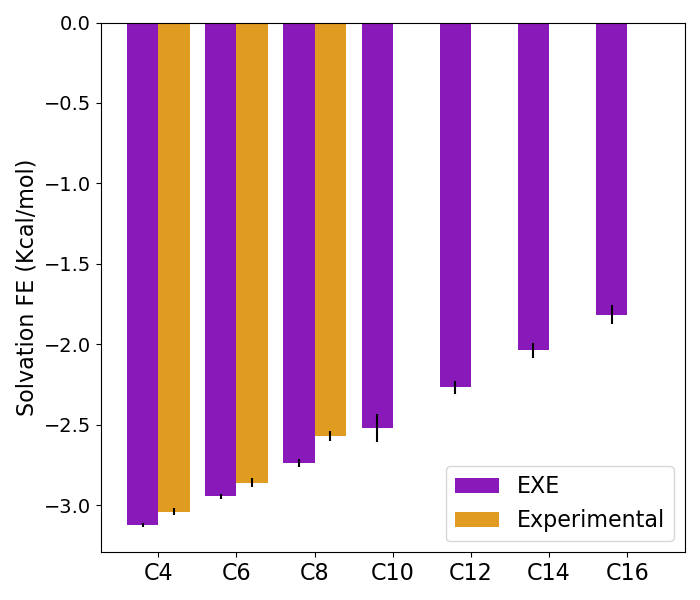}
    \caption{We validated the initial parameterization of our small molecules by comparing the absolute solvation free energy of aldehydes or varying length to available experimental data. We observed that the absolute solvation free energy for acyl chain lengths C4, C6, and C8 are within 0.2 kcal/mol of the experimental measurements. Additionally the trend in solvation free energy as the chain lengths extend is consistent with experimental data. This allows us to confidently move on with the relative free energy calculations using this same parameterization.}
    \label{SI:acyl_solv}
\end{figure}

We first compute the absolute solvation free energy of each aldehyde chain from 4 to 16 carbons and compare against available experimental data in order to ensure there are no significant issues affecting overall molecular behavior with the paramaterization of our molecules or other common simulation settings common to all simulations. Where experimental data was available, our solvation free estimates properly followed the trend in solvation free energy, though they are not necessarily expected to be within experimental error (Figure \ref{SI:acyl_solv}) allowing us to move on to running the relative solvation free energy simulations using both the standard expanded ensemble (EXE) method as well as our MT-REXEE method.

\begin{figure}
    \centering
    \includegraphics[width=1.0\linewidth]{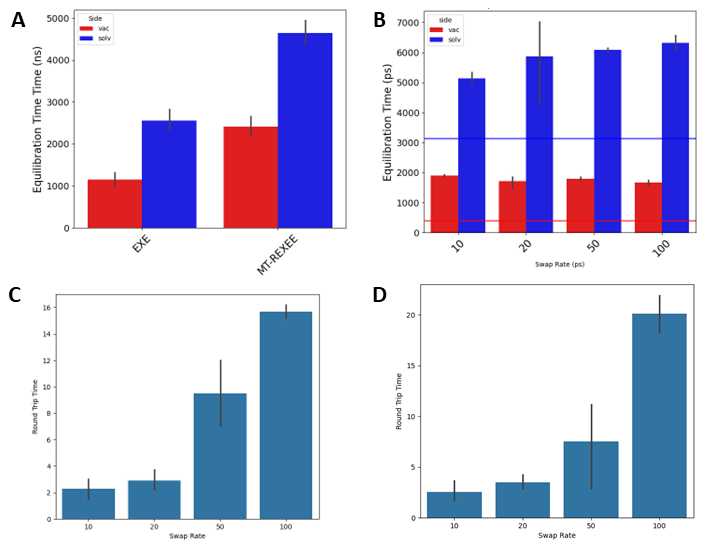}
    \caption{We optimized the performance of the MT-REXEE method by minimizing the equilibration time and the global round trip time. First we evaluated a base method using our acyl test system and determined that when using 3X as many states for MT-REXEE (due to redundant end states) and a swapping frequency of 20 ps we can achieve Wang-Landau weight equilibration times ~2X longer than that of EXE method alone (A). When we look at the solvation free energy of the MUP1 ligands we see the same general trend in Wang-Landau weight equilibration times between MT-REXEE and EXE with no dependence on the swapping frequency (B). We then evaluated the global round trip time for the MUP1 system for both the vacuum (C) and solvent (D) MT-REXEE simulations and showed that generally as the swapping frequency decreases (thus swaps occur more frequently) the round trip time decreases. However, once the swapping frequency reaches $<$20 ps we observe that simulations can become stuck swapping between the same two simulations multiple times thus extending the round trip time. The optimal swapping frequency is likely system dependent and thus this rate may be a good starting point, but it should be optimized for each system.}
    \label{SI:equil_and_opt}
\end{figure}

We utilize the same $\lambda$ state spacing for both the EXE and MT-REXEE simulations with the exception of the inclusion of redundant end states (3 for vacuum simulations and 5 for solvent simulations) for the MT-REXEE simulations. These redundant end states do lead to a slight increase in the time needed to optimize the $\lambda$ weights. This translates to mean vacuum weight equilibration times of $1.16 \pm 0.34$ ns and $2.41 \pm 0.51$ ns for EXE and MT-REXEE respectively and solvent weight equilibration times of $2.56 \pm 0.56$ and $4.64 \pm 0.65$ for EXE and MT-REXEE respectively (Figure~\ref{SI:equil_and_opt}A). The MT-REXEE simulations have 3X as many $\lambda$ states so increasing the weight equilibration time by $\sim$2X is reasonable. In order to directly compare the two methods, the length of the simulation after weight equilibration was 20 ns for vacuum simulations and 30 ns for solvent simulations using both EXE and MT-REXEE. 

As intended, there is not a statistically significant difference between the relative free energy estimate obtained from EXE and MT-REXEE (Figure \ref{fig:aldehyde_summary} B and \ref{SI:FE_scatter}A). Using equation \ref{eq:thermo_MTREXEE}, we obtain the error likely introduced through the inclusion of configurational swapping steps in MT-REXEE and obtain values which are between $0.0079 \pm 0.134$ kcal/mol and $0.025 \pm 0.052$ kcal/mol (Figure \ref{fig:aldehyde_summary}C). This error is within the statistical error for every one of these measurements which validates that, for this simple system, the MT-REXEE method does not introduce significant error to the relative free energy estimate. 

We need to ensure that the limited error introduced is not due to the simulations becoming trapped in an individual replica as this would essentially become equivalent to a stand-alone EXE simulation. This can be demonstrated through the comparison of round-trip times for the replica swaps, that is, the time it takes for the configuration to sample all other replicas and return to its initial replica. This round trip time was determine to be $2.49 \pm 1.04$ and $4.00 \pm 1.54$ for vacuum and solvent simulations respectively (Figure \ref{fig:aldehyde_summary}D). This means that a minimum of two round trips occur per replica in the simulations conducted here with a 20 ps swapping frequency. We will explore later in the manuscript how varying the rate of this swapping frequency also varies the round trip time for the simulations.

\subsection{MUP1 Ligands System Validation}
\begin{figure}
    \centering
    \includegraphics[width=0.9\linewidth]{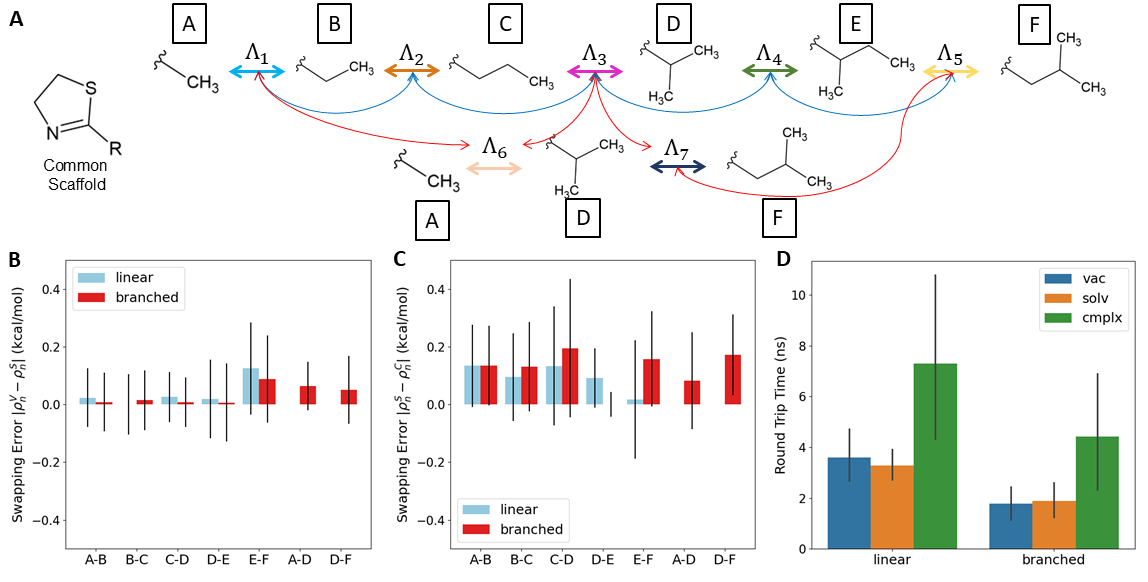}
    \caption{(A) The swapping network defines the relative transformations occurring in both the linear (blue) and branched (red) networks for both the relative solvation free energy and relative binding free energy calculations. (B) 
    The additional error in relative solvation free energy calculations introduced using the MT-REXEE method with both the linear and branched paths is statistically indistinguishable from zero. This supports our claim that any error introduced by the conformational swaps is equivalent in solvent and vacuum and will cancel when the relative free energy is computed. (C) This same cancellation or errors also occurs for solvent and complex simulations used to compute the relative binding free energy. (D) The introduction of the branched swapping pattern has the same reduction of round trip time effect on the vacuum, solvent, and complex simulations though the round trip time was shown to significantly increase in complex simulations.}
    \label{fig:MUP1_solv_summary}
\end{figure}

\begin{figure}
    \centering
    \includegraphics[width=0.9\linewidth]{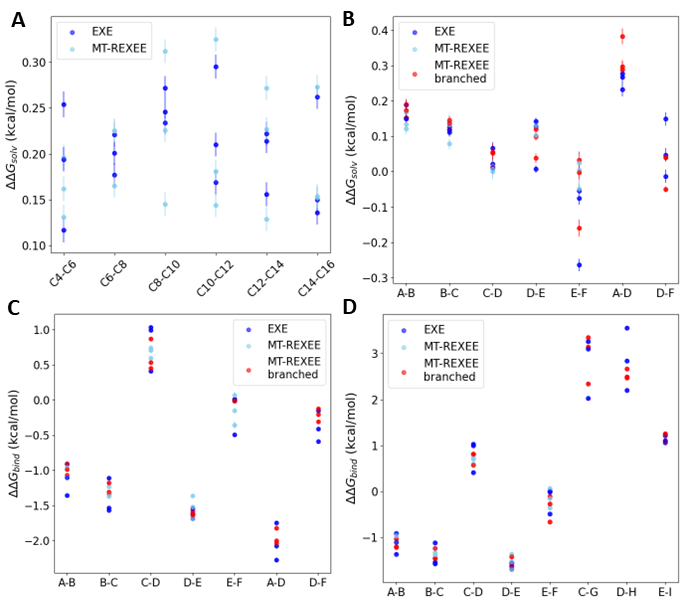}
    \caption{For each transformation we plot the three replicas relative free energy estimate using both MT-REXEE and standard EXE. The error bars represent the MBAR error approximated using pymbar for each individual simulation. We show the solvation free energy for the acyl chain growth (A) and MUP1 ligands as well as the binding free energy for the MUP1 ligands using the standard set (C) and the set expanded to include diverse heavy atoms (D). None of the systems evaluated for relative solvation free energy or relative binding free energy show any indication of systematic error introduced from the MT-REXEE method.}
    \label{SI:FE_scatter}
\end{figure}

We again began with computing the solvation free energy of the ligands to validate that the MT-REXEE method does not affect cycle closure of the relative free energy measurements. However, for this system in addition to computing a straight path of adding functional groups to the molecule, we also added alternate transformations to create a branched swapping network (Figure \ref{fig:MUP1_solv_summary}A and \ref{SI:FE_scatter}B--C). These calculations validate that increasing the complexity of the swapping network does not introduce significant additional error to the system. This helps demonstrate the generalization of this method to diverse systems for which complicated branched transformation maps may be necessary. The only limitation to swapping between different transformations is that one end state must be shared between the two replicates.

As with the chain growth system we compute the relative free energy of each transformation using the absolute free energy of the end states as well as the relative transformation using standard EXE to validate our MT-REXEE calculations. When comparing the AFE and standard EXE to both the linear path MT-REXEE and the branched path, we see no significant difference in relative solvation free energy (Figure \ref{fig:MUP1_solv_summary}B). Using the same transformation network, we also computed relative binding free energy values using both EXE and MT-REXEE methods. The difference between the EXE method and both the branched and linear transformation paths were found to be within simulation error (Figure \ref{fig:MUP1_solv_summary}C). These calculations demonstrate that as long as the sampling is sufficient in the simulation the MT-REXEE method will produce the same relative free energy estimate value as EXE alone. We can also demonstrate that the introduction of these additional swapping paths has a significant reduction on the round trip time due to the additional swapping paths which were introduced (Figure \ref{fig:MUP1_solv_summary}D). The significant increase in the round trip time in complex simulations was found to not be due to swaps being rejected, but rather because the complex simulations were found to be more prone to becoming temporarily stuck swapping between several $\lambda$ states within a given simulation, reducing the likelihood of two simulations visiting compatible $\lambda$ states at the end of a given simulation. This effect can be reduced by lowering the Wang-Landau threshold for weight equilibration though the uneven weight sampling was seen only during short time frames (5-10 ps) and over longer time frames (100 ps) the state sampling was even across $\lambda$ states with the equilibration threshold used in this study.

\begin{figure}
    \centering
    \includegraphics[width=0.9\linewidth]{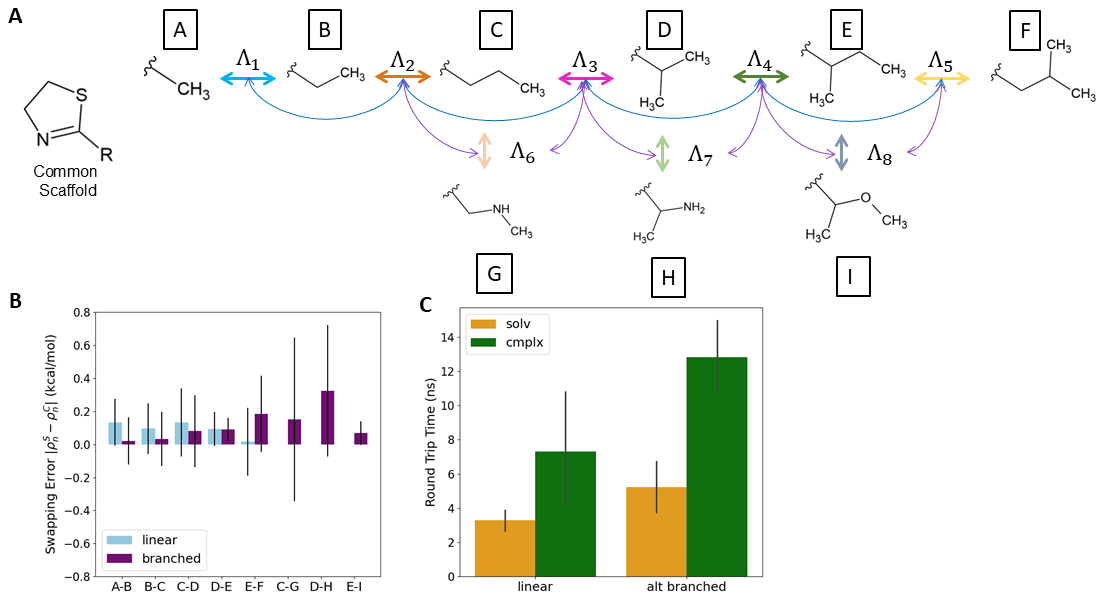}
    \caption{(A) The swapping network shown here introduces diversity in the heavy atoms involved in the relative transformations for both the complex and solvent simulations. (B) Despite this additional complexity, there is again no statistically significant error introduced to the computation of the relative binding free energy for any of the transformations. (C) The introduction of these additional relative transformations does increase the round trip time by 60\% which is consistent with the 50\% increase in the end states which must be sampled to complete a round trip in this swapping network.}
    \label{fig:MUP1_b2}
\end{figure}

All relative solvation and binding free energy calculation which have been completed thus far involve dummy hydrogen and carbon atoms only. To ensure that the method is generalizable to a wide range of potential relative transformations, we included additional transformations which involved altering a heavy atom (in this case carbon) to another heavy atom (in this case oxygen or nitrogen) (Figure \ref{fig:MUP1_b2}A and \ref{SI:FE_scatter}D). The introduction of this diversity in heavy atoms involved in the transformations resulted in no significant error introduced into the free energy estimate (Figure \ref{fig:MUP1_b2}B). This branched network was shown to significantly increase the round trip times as there are three additional end states which must be visited in order to complete a round trip (Figure \ref{fig:MUP1_b2}C). This increase to the round trip time is approximately linearly related to the number of unique end states in the system, which is as expected given that only a given simulation can only engage in one conformational swap in each iteration. As we continue to increase the complexity of the transformation network by introducing additional unique end states, we also increase the importance of planning the swapping network to minimize the round trip time.

\section{Discussion}
The MT-REXEE method allows for the computation of multiple relative free energies when there is a common scaffold for the transformations. In this study we demonstrate that the conformational exchanges which occur between replicas in order to enhance sampling for all calculations to not introduce error into the free energy calculation in the event that sufficient sampling is completed. We have shown that this is the case for both linear and branched swapping patterns as well as relative transformations which involve a variety of heavy atoms (carbon, oxygen, and nitrogen). 

The primary limitation of the current method is that this implementation requires redundant end states which increases the Wang-Landau weight equilibration time and also increases the time it takes for free energy estimates to converge, due to decreased sampling of intermediate states. The computational cost of running MT-REXEE does scale linearly with the number of transformations being computed, the same as any other large scale scan of multiple alchemical mutations.  The only difference is that some level of exchange between simulation tasks which can be carried out relatively infrequently compared to replica exchange is needed. For binding free energy calculations, the MT-REXEE method also samples separate protein configurations with each alchemical transformation, and this sharing between alchemical states may be crucial to getting accurate estimates for flexible protein systems. This is a clear advantage over $\lambda$-dynamics, which is significantly faster, but only samples one protein configurational ensemble~\cite{robo_fast_2023}.

Further work will focus on demonstrating the benefit of this enhanced conformational sampling in larger systems,  which could significantly benefit using MT-REXEE over other methods. The application of this method is most well suited for systems which feature large flexible proteins or ligands. This is particularly relevant for the rise of interest in study of peptide drugs which are not well-suited for traditional methods due to their flexibility. 

Though drug discovery is the primary focus of most RFE studies, the application of these calculations can extend far beyond small molecule and peptide drugs. For example, it could be applied to enzyme engineering problems which involves protein--protein binding interfaces, which often involve significant rearrangement. 
%The full benefit of this additional conformational sampling will be demonstrated as well extend to complicated systems like the relative binding free energy of the acyl carrier protein with varying length carbon chains to the $\beta$-ketoacyl reductase, FabG which features both a inter protein interface as well as a flexible substrate. 

\clearpage
\bibliography{citations/AFE_intro, citations/applications, citations/enhanced_sampling, citations/Software}
\clearpage

\end{document}